\documentclass[twocolumn,showpacs,preprintnumbers,amsmath,amssymb]{revtex4}
\usepackage{amssymb}%math symbols
\usepackage{graphicx}% Include figure files
\usepackage{dcolumn}% Align table columns on decimal point
\usepackage{bm}% bold math

\begin{document}
\title[]{Electrically-induced polarization selection rules of a graphene quantum dot}

\author{Qing-Rui Dong}
\address{School of Physics and Electronics, Shandong Normal University, Jinan, Shandong, 250014, People's Republic of China}
\author{Yan Li}
\address{School of Physics and Electronics, Shandong Normal University, Jinan, Shandong, 250014, People's Republic of China}
\author{Chun-Xiang Liu}
\address{School of Physics and Electronics, Shandong Normal University, Jinan, Shandong, 250014, People's Republic of China}

\begin{abstract}
We study theoretically the single-electron triangular zigzag graphene quantum dot in uniform in-plane electric fields.
The far-infrared absorption spectra of the dot are calculated by the tight-binding method.
The energy spectra and the distribution of wave functions are also presented to analyse the far-infrared spectra.
The orthogonal zero-energy eigenstates are arranged along to the direction of the external field.
The remarkable result is that all intraband transitions and some interband transitions are forbidden when the absorbed light is polarized along the direction of the electric field.
With $x$-direction electric field,  all intraband absorption is $y$ polarized due to the electric-field-direction-polarization selection rule.
Moreover, with $y$-direction electric field, all absorption is either $x$ or $y$ polarized due to the parity selection rule as well as to the electric-field-direction-polarization selection rule.
Our calculation shows that the formation of the FIR spectra is co-decided by the polarization selection rules and the overlap between the eigenstates of the transition.
\end{abstract}

\maketitle

\section{Introduction}
Graphene, a single layer of carbon atoms arranged in a two-dimensional honeycomb lattice, was first successfully fabricated in 2004\cite{Novoselov04}. Due to the exceptional properties, graphene has attracted enormous research interest and exhibited great application potential in next-generation electronics\cite{Novoselov12} and optoelectronics\cite{Xia14}. Much of the current understanding of the electronic properties of graphene has
been reviewed by Castro-Neto\cite{neto2009electronic}, transport properties by Das Sarma\cite{sarma2011electronic} and many-body effects by Kotov\cite{kotov2012electron}.
However, a gap has to be induced in the gapless graphene for its real applications in electronic devices.
For this purpose, graphene quantum dots (GQDs) have been proposed as one of the most promising kinds of graphene nanostructures\cite{gucclu2014graphene}.
GQDs exhibit the unique electronic, spin and optical properties, which allow them hold great application potential in electronics and optoelectronics such as super capacitor\cite{liu2013superior}, flash memory\cite{joo2014graphene}, photodetector\cite{kim2014high} and phototransistor\cite{konstantatos2012hybrid}.
On the other hand, with recent developments of fabrication techniques, it is possible to cut accurately the bulk graphene into different sizes and shapes, such as hexagonal zigzag quantum dots, hexagonal armchair quantum dots, triangular zigzag quantum dots and triangular armchair quantum dots\cite{bacon2014graphene}.

Further applications of GQDs require a thorough knowledge of their electronic properties.
The electronic and magnetic properties of GQDs depend strongly on their shapes and edges\cite{Gu09,abergel2010properties,Potasz12,jiang2017energy}.
Moreover, for zigzag GQDs, especially triangular GQDs (TGQDs), there appears a shell of degenerate states at the Dirac points and the degeneracy is proportional to the edge size\cite{potasz2010zero,JZT2014}.
The electronic states of TGQDs can be classified by the group theory according to irreducible representations of the $C3$ symmetry group\cite{gucclu2014graphene}.
As a result of the degenerate zero-energy band, magnetism arising in graphene nanostructures (nanoflakes, quantum dots and nanoribbons) has recently collected rich literature\cite{sun2017magnetism,basak2016optical,hawrylak2016carbononics}.
The key feature for device application of GQDs is the ability to manipulate their electronic structures.
Therefore, one of the flourishing fields of exploration is the influence of external fields on the degenerate zero-energy band.
The electronic structure and magnetization relating to the zero-energy band can be manipulated electrically\cite{Chen10,Ma12PRB,Dong13,farghadan2014Iran,dong2014electronic,szalowski2017ferrimagnetic}, optically\cite{Gu13} and magnetically\cite{szalowski2015Poland}.
Moreover, the effect of an external magnetic field on electron-hole interactions in GQDs has been explored.\cite{Peeters2017exciton}.
In particular, the electrical manipulation of the zero-energy band of such GQDs is quite important for the operation of related devices, since it is easier to generate the potential field through local gate electrodes than the optical or magnetic field.
The advantage of applying external electric fields is that these fields can adjust the splitting of the degenerate zero-energy band and then can adjust the optical transition wavelength.
However, it is rather rare to study the influence of electric fields on the optical properties relating to the zero-energy band\cite{abdelsalam2016electro,dong2017optical}.

In this paper, we concentrate on the effects of two uniform in-plane electric fields on the far-infrared (FIR) absorption spectra of a TGQD.
The energy spectra and the distribution of wave functions are also presented to analyse the FIR spectra.
The remarkable result is that all intraband transitions and some interband transitions are forbidden when the absorbed light is polarized along the direction of the electric field.
Our calculations show that the formation of the FIR spectra is co-decided by the polarization selection rule and the overlap between the eigenstates of the transition.

\section{The electric fields and the energy spectra}
\begin{figure}[htp]
\centering
\includegraphics[height=12cm]{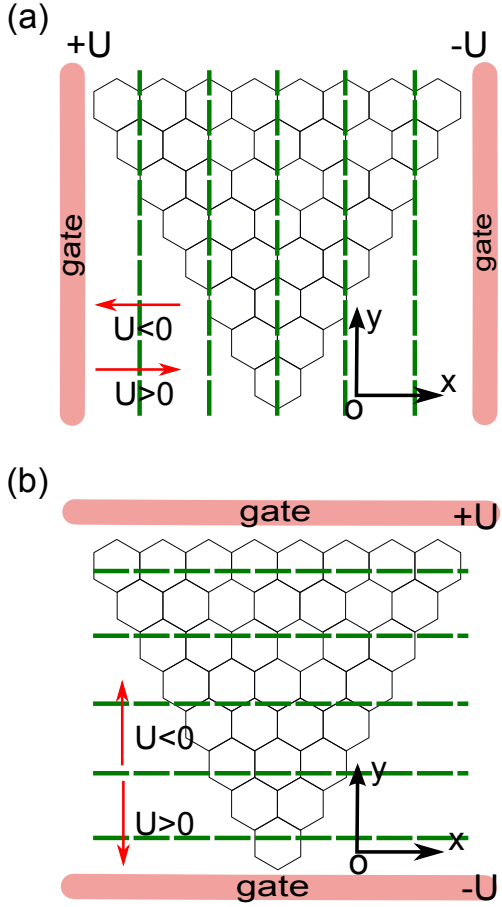}\caption{\label{fig:Fig1}
The uniform electric field (a) $EF1$ and (b) $EF2$ applied to a TGQD ($N_{s}$ = 8).
The TGQD is assumed to lie in the $x-y$ plane, where an edge of the TGQD is along the $x$ axis and the perpendicular bisector is along the $y$ axis.
Each electric field is generated by two parallel electrodes with electrostatic potentials $\pm U$.
The contours of the electrostatic potentials are shown (green dashed curves) and the directions of the electric fields are indicated (red arrows).}
\end{figure}

In Fig. \ref{fig:Fig1}, two typical in-plane uniform electric fields are applied respectively to a TGQD with the size $N_s$ = 8, where $N_s$ is the number of carbon atoms in each side of the dot.
Each electric field is generated by two parallel gate electrodes with opposite electrostatic potentials $\pm U$.
In Fig. \ref{fig:Fig1}(a), the direction of the electric field $EF1$ is along the $x$ axis, and the electrostatic potentials of the left and right half of the dot have opposite signs.
In Fig. \ref{fig:Fig1}(b), the direction of the electric field $EF2$ is along the $y$ axis, and the electrostatic potentials of the left and right half of the dot have same signs.
From the point of view of symmetry, the electric field $EF1$ destroys the mirror symmetry of the system, where the plane of symmetry is the vertical plane containing the perpendicular bisector along the $y$ axis.
Unlike $EF1$, the electric field $EF2$ retains the mirror symmetry.
Later it will be seen that this difference in symmetry has important consequences for the FIR spectra.

The low-energy electronic structure of a GQD subjected to an in-plane electric field can be calculated by means of the tight-binding method\cite{Chen10,JZT2014}.
In the low-energy range, the tight-binding Hamiltonian with the nearest-neighbor approximation proves to give the same accuracy as first-principle calculations\cite{Abergel10}.
The Hamiltonian equation of the system is $H|\Psi_{j}(\textbf{r})\rangle=E_{j}|\Psi_{j}(\textbf{r})\rangle$ and the tight-binding Hamiltonian with the nearest-neighbor approximation is\cite{Ma12,jiang2017comparative}
\begin{equation}\label{eqn:1}
H=\sum_{n}{(\varepsilon_{n}+U_{n})C^{+}_{n}C_{n}}+\sum_{<n,m>}{t_{n,m}C^{+}_{n}C_{m}},
\end{equation}
where $n$, $m$ denote the sites of carbon atoms in graphene, $\varepsilon_{n}$ is the on-site energy of the site $n$, $U_{n}$ is the electrostatic potential of the site $n$ obtained by solving a Laplace equation, $t_{n,m}$ is the hopping energy and $C^{+}_{n}$ ($C_{n}$) is the creation (annihilation) operator of an electron at the site $n$. The summation $<n,m>$ is taken over all nearest neighboring sites.
Due to the homogeneous geometrical configuration, the on-site energy and the hopping energy may be taken as $\varepsilon_{n}$ = 0 and $t_{n,m}$ = 2.7 eV.

\begin{figure}[htp]
\centering
\includegraphics[height=7cm]{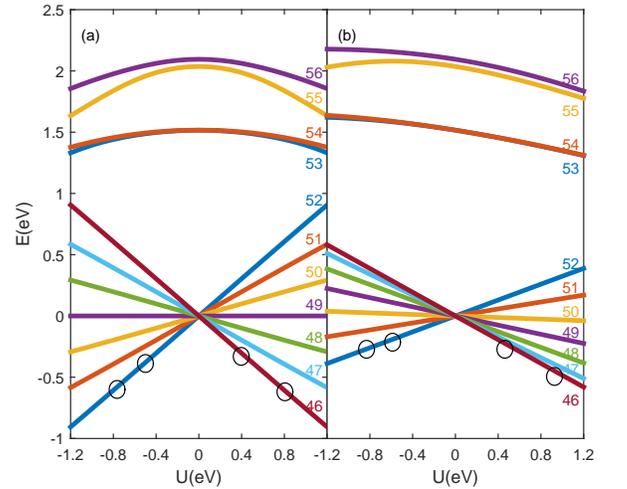}\caption{\label{fig:Fig2}
The single-electron energy spectra of the TGQD ($N_{s}$ = 8) with (a) $EF1$ and (b) $EF2$.
Each level is labeled by eigenstate index and is shown in a different color.
The open circles denote the ground-state level of the transition.}
\end{figure}

Fig. \ref{fig:Fig2} shows the single-electron energy spectra of the TGQD ($N_{s}$ = 8) with $EF1$ and $EF2$.
With $EF1$, the energy spectrum for $\pm U$ is bilaterally symmetrical because the TGQD is bilaterally symmetrical.
With $EF2$, the energy spectrum for $\pm U$ is bilaterally asymmetrical because the TGQD is not up-down symmetrical.
The TGQD contains $97$ atoms and thus there are $97$ eigenstates $\Psi_{j}(\textbf{r})$.
For the study of the single-electron FIR spectra, it is enough that we consider only $11$ levels around the Fermi level.
Seven of these eigenstates (eigenstate index: 46-52) are belong to the zero-energy band.
The other four eigenstates (eigenstate index: 53-56) possess higher energy.
In Fig. \ref{fig:Fig2}, the ground-state level of the transition is indicated.
According to the ground-state level and the chemical potential of the leads, one can guarantee that there exists only one electron in the dot.\cite{dong2017optical}

\begin{figure*}[htp]
\centering
\includegraphics[height=6.5cm]{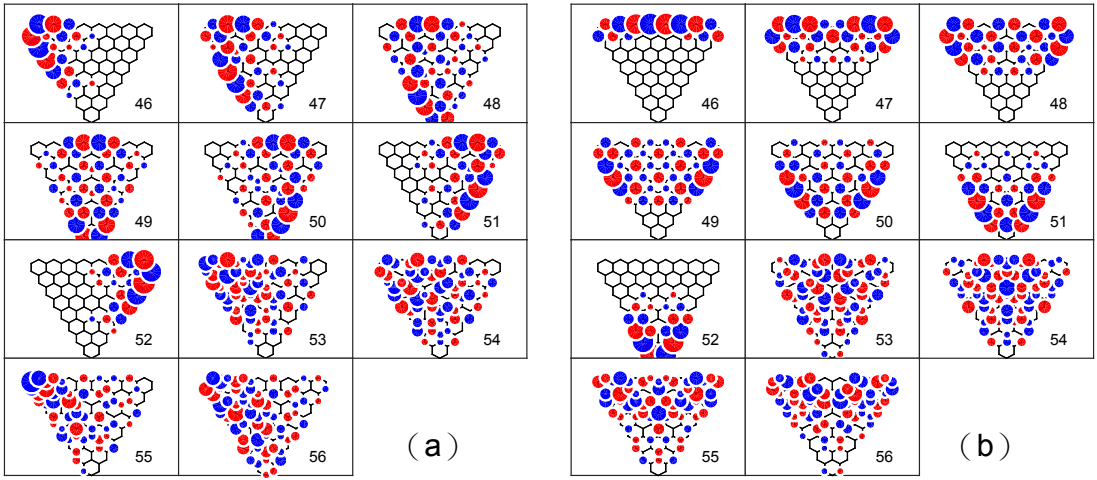}\caption{\label{fig:Fig3}
The amplitude $\omega_{ji}$ for all states in the TGQD ($N_{s}$ = 8) with (a) $EF1$ ($U=0.5$ eV) and (b) $EF2$ ($U=0.5$ eV), where the red (blue) circles denote the amplitudes $\omega_{ji}$ that are positive (negative).
The amplitude is proportional to the radius of a circle.
All eigenstate indexes are corresponding to those indexes in Fig. \ref{fig:Fig2}.}
\end{figure*}

We also give the distribution of wave functions, which is helpful to discuss the calculated FIR spectra later.
A wave function is represented as
\begin{equation}\label{eqn:1x}
\Psi_{j}(\textbf{r})=\sum_{i}\omega_{ji}\varphi_{i}(\textbf{r}),
\end{equation}
where $\varphi_{i}(\textbf{r})$ is the Wannier function localized at the site $i$ and $\omega_{ji}$ is the amplitude for $\Psi_{j}(\textbf{r})$.
The operator $C_n$ in the Hamiltonian (\ref{eqn:1}) annihilates an electron in the state described by the Wannier function $\varphi_{n}(x)$.
We are able to calculate the amplitude $\omega_{ji}$ for $\Psi_{j}(\textbf{r})$ and all of them are found to be real.
Fig. \ref{fig:Fig3} shows the distribution of the wave functions with $EF1$ and $EF2$, respectively.

The eigenstates of the zero-energy band are almost constant with $U$ while the eigenstates of the nonzero-energy band are mixed continuously with $U$\cite{dong2014electronic}.
With $EF1$, the orthogonal zero-energy states (46-52) are arranged from left to right in the dot.
With $EF2$, the orthogonal zero-energy states (46-52) are arranged from top to bottom in the dot and all eigenstates are symmetrical or antisymmetrical due to the mirror symmetry.
%One of the wave functions is entirely localized on edge sites for a TGQD with N=odd and there are no such wave functions for a TGQD with N=even.\cite{Ezawa07}

\section{The unpolarized FIR spectra}
Using the Fermi golden rule with the electric-dipole approximation for the perturbing unpolarized light, the transition probability from the ground state to the $l$th excited state can be calculated as\cite{dong2007,abdelsalam2016electro}
\begin{equation}\label{eqn:2}
A_l\varpropto|\langle \Psi_{l}|\textbf{r}|\Psi_{0}\rangle
|^{2}\delta(E_{l}-E_{0}-\hbar \omega),
\end{equation}
$\Psi_{0}$ is the ground state of the zero band and the corresponding level is indicated in Fig. \ref{fig:Fig2}.

\begin{figure}[htp]
\centering
\includegraphics[width=7.5cm]{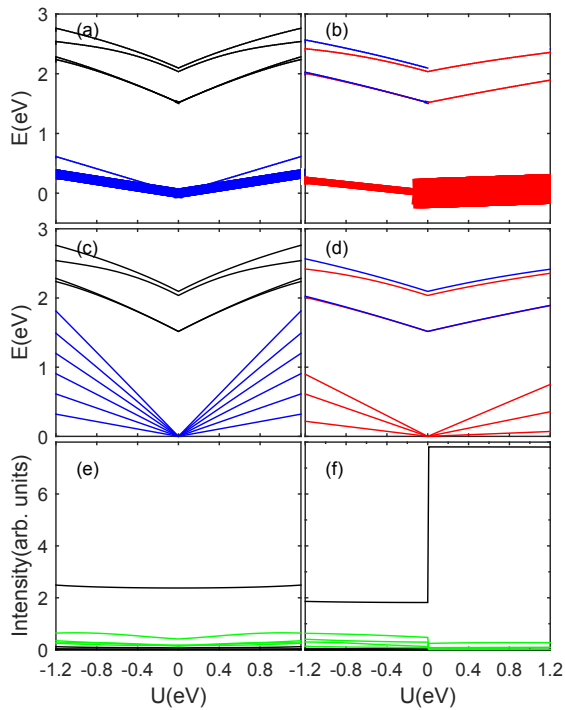}\caption{\label{fig:Fig4}
The calculated FIR spectra of the single-electron TGQD ($N_{s}$ = 8).  (a) The practical FIR spectra with $EF1$. (b) The practical FIR spectra with $EF2$. (c) The fine FIR spectra with $EF1$. (d) The fine FIR spectra with $EF2$.
In (a)-(d), spectra show the energy of absorbed light as a function of electric field and red (blue) lines indicate $x$-polarized ($y$-polarized) absorptions.
(e) The peak intensities with $EF1$. (f) The peak intensities with $EF2$. In (e) and (f),  black (green) lines indicate intraband (interband) peaks.}
\end{figure}

Fig. \ref{fig:Fig4} shows the calculated FIR spectra of the single-electron TGQD ($N_{s}$ = 8) with $EF1$ and $EF2$ respectively.
In Fig. \ref{fig:Fig4}(a-b), we have included only the transitions which have a peak intensity of more than 1\% of the maximum value.
By excluding those absorption peaks with a relatively low intensity, the FIR spectra tend to simulate experimental observation and thus are more practical.
In Fig. \ref{fig:Fig4}(c-d), we have included all peaks with a nonzero intensity, which helps to identify whether the selection rule exists.
Thus, Fig. \ref{fig:Fig4}(c-d) can been regarded as the fine structure of the FIR spectra.
To make a distinction, we call Fig. \ref{fig:Fig4}(a-b) as the practical FIR spectra and call Fig. \ref{fig:Fig4}(c-d) as the fine FIR spectra.
To show exactly the difference among the peak intensities, we have plotted them with $EF1$ and $EF2$ in Fig. \ref{fig:Fig4}(e-f).
As a general feature of the calculated spectra shown in Fig. \ref{fig:Fig4}(a-d), one can see that each spectrum has two branches as a major component, where the higher one comes from the interband transitions and the lower from the intraband transitions.

Now we discuss the effect of $EF1$ on the FIR spectra.
In the Fig. \ref{fig:Fig4}(a), (c) and (e), the FIR spectra are bilaterally symmetrical because the energy spectrum with $EF1$ is bilaterally symmetrical, although the ground eigenstate of the transition on $U>0$ is different from that on $U<0$.
For example, the ground state is the eigenstate $46$ on $U>0$ and the eigenstate $52$ on $U<0$.
The practical FIR spectrum in Fig. \ref{fig:Fig4}(a) shows that two intraband peaks appear while the other intraband peaks disappear.
However, the fine FIR spectrum in Fig. \ref{fig:Fig4}(c) shows that there are not real forbidden transitions.
Fig. \ref{fig:Fig4}(e) shows that the intensity of the second intraband peak is about 5\% of the intensity of the strongest intraband peak.
Four interband peaks possess higher intensity than the second intraband peak.
Both the evolution from Fig. \ref{fig:Fig4}(c) to Fig. \ref{fig:Fig4}(a) and the difference among the peak intensities can be explained by the overlap between the initial and final eigenstate of the transition [Fig. \ref{fig:Fig3}(a)].
For example, at $U=0.5$ eV, the initial eigenstate $46$ is located in the left-top part of the dot.
Comparing with the initial eigenstate $46$, the final eigenstate $47$ is moved slightly toward the right while the final eigenstate $48$ is moved further toward the right.
As the final eigenstate changes from the eigenstate $47$ to the eigenstate $52$, the overlap between the initial and final eigenstate becomes smaller and smaller.
When the final eigenstate is the eigenstate $49$, the absorption peak disappears in the practical FIR spectra.

In the following, we discuss the effect of $EF2$ on the FIR spectra.
In Fig. \ref{fig:Fig4}(b), (d) and (f), the FIR spectra are not bilaterally symmetrical because the energy spectrum with $EF2$ is not bilaterally symmetrical.
The fine FIR spectrum in Fig. \ref{fig:Fig4}(d) shows that there are only three intraband peaks, which implies that there exist real forbidden transitions.
The selection rule will be detailedly discussed later.
In the practical FIR spectra [Fig. \ref{fig:Fig4}(b)], there appears only one intraband peak and the peak intensity on $U>0$ is much higher than the peak intensity on $U<0$.
Moreover, there are two interband peaks on $U>0$ and four interband peaks on $U<0$.
Both the evolution from Fig. \ref{fig:Fig4}(d) to Fig. \ref{fig:Fig4}(b) and the difference among the peak intensities can be explained by the overlap between the initial and final eigenstate of the transition [Fig. \ref{fig:Fig3}(b)].

The formation of the practical FIR spectra is closely related to the overlap between the initial and final eigenstate.
In the next section, it can be shown that polarization selective rules also affect the formation of the practical FIR spectra.

\section{The polarized spectra with $X$-direction electric field}

\begin{figure*}[htp]
\centering
\includegraphics[width=16cm]{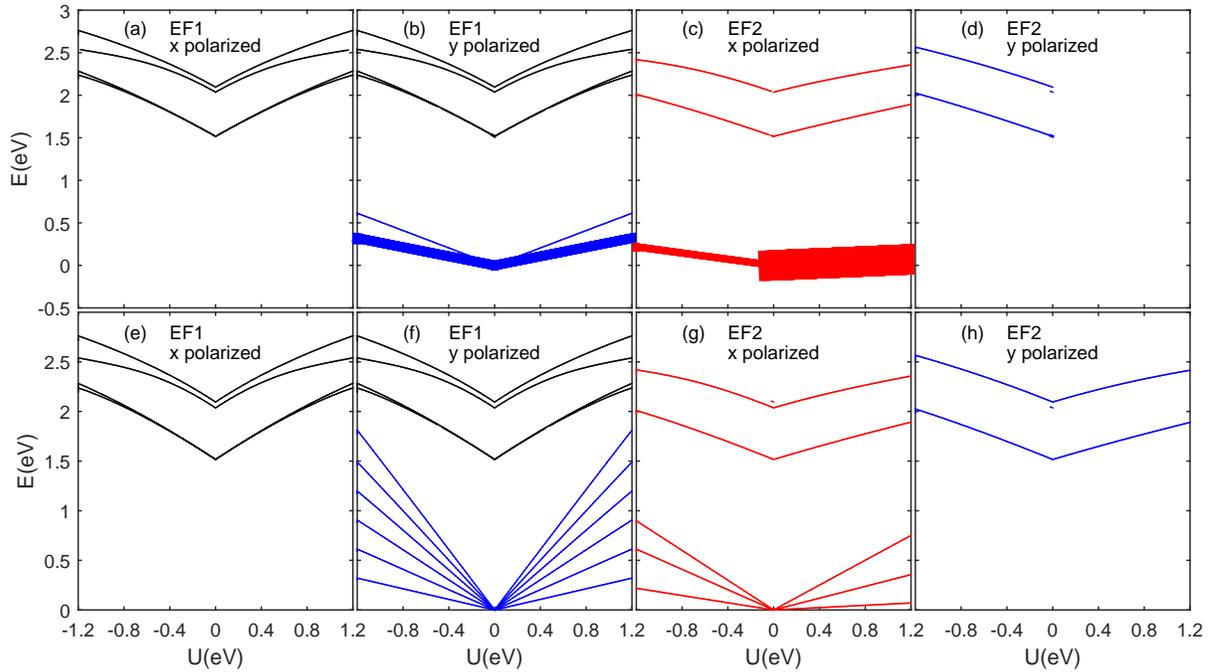}\caption{\label{fig:Fig5}
The polarized FIR spectra of the TGQD ($N_{s}$ = 8). Spectra show the energy of absorbed light as a function of electric field.
(a)-(d) The practical FIR spectra including only the transitions which have a peak intensity of more than 1\% of the maximum value.
In (a)-(d), the width of each line is roughly proportional to the peak intensity.
(e)-(h) The fine FIR spectra including all peaks with nonzero intensities.
In each subplot, the electric field and the polarization of incident light are labeled.
Red (blue) lines indicate that the peaks only appear in the $x$-polarized ($y$-polarized) FIR spectra.}
\end{figure*}

The analysis of polarized spectra is helpful to understand the formation of the FIR spectra in more detail.
One selected spectrum can be decomposed to $x$ and $y$ polarization,
\begin{equation}\label{eqn:3}
\left\{ \begin{gathered}
  A_l^x\varpropto|\langle \Psi_{l}|x|\Psi_{0}\rangle
|^{2}\delta(E_{l}-E_{0}-\hbar \omega) \\
  A_l^y\varpropto|\langle \Psi_{l}|y|\Psi_{0}\rangle
|^{2}\delta(E_{l}-E_{0}-\hbar \omega)  \\
\end{gathered}  \right.
\end{equation}
According to the irreducible theory of the symmetry group\cite{elliott1979symmetry}, symmetry leads to selection rules or forbidden transitions.
To a certain system, the transition probability for the polarized light $A_l^x$ or $A_l^y$ is a component of $A_l$.
Thus, the polarization may cause the decrease of the transition probabilities.
Moreover, the $x$ and $y$ polarization may cause forbidden transitions due to the symmetry in $x$ or $y$ direction.
In the section, we analyze the formation of the polarized practical FIR spectra with $EF1$.
With $EF1$, the $x$- and $y$-polarized practical FIR spectrum are shown in Fig. \ref{fig:Fig5}(a) and (b), respectively.
With $EF1$, the $x$- and $y$-polarized fine FIR spectrum are shown in Fig. \ref{fig:Fig5}(e) and (f), respectively.

Although there are not forbidden transitions in the unpolarized spectrum [Fig. \ref{fig:Fig4}(c)], the intraband transitions are forbidden in the $x$-polarized fine FIR spectrum [Fig. \ref{fig:Fig5}(e)].
Transition selection rules relating to electrical dipole radiation are usually classified into two categories: parity selection rules and angular momentum selection rules.
Obviously, Fig. \ref{fig:Fig3} suggests that the polarization selection rule shown by the $x$-polarized fine FIR spectrum is not belong to the two categories.
The direction of the electric field $EF1$ is along the $x$ axis.
That is to say, all intraband transitions are forbidden when the absorbed light is polarized along the direction of the electric field.
Thus, in the practical FIR spectrum [Fig. \ref{fig:Fig5}(a)], the intraband absorption peaks disappear due to the electric-field-direction-polarization selection rule.

Fig. \ref{fig:Fig5}(f) shows that there are not forbidden transitions in the $y$-polarized fine FIR spectrum.
Meanwhile, four intraband absorption peaks disappear in the $y$-polarized practical FIR spectrum [Fig. \ref{fig:Fig5}(b)].
Thus, in the evolution from Fig. \ref{fig:Fig5}(f) to Fig. \ref{fig:Fig5}(b), the reason why four intraband absorption peaks disappear is too little overlap between the initial and final eigenstate of the transition.

Fig. \ref{fig:Fig4}(c) can be regarded as the combination of Fig. \ref{fig:Fig5}(e) and (f).
Meanwhile, Fig. \ref{fig:Fig4}(a) can be regarded as the combination of Fig. \ref{fig:Fig5}(a) and (b).
Since the $x$-polarized intraband transitions are forbidden, all intraband absorptions are $y$ polarized in the unpolarized FIR [Fig. \ref{fig:Fig4}(a) and (c)].

\section{The polarized spectra with $Y$-direction electric field}

In the section, we analyze the formation of the practical polarized FIR spectra with $EF2$.
With $EF2$, the $x$- and $y$-polarized practical FIR spectrum are shown in Fig. \ref{fig:Fig5}(c) and (d), respectively.
With $EF2$, the $x$- and $y$-polarized fine FIR spectrum are shown in Fig. \ref{fig:Fig5}(g) and (h), respectively.

In the $x$-polarized fine FIR spectrum [Fig. \ref{fig:Fig5}(g)], half of the transitions are forbidden due to a parity selection rule.
For example, the transition from the eigenstate $46$ to the eigenstate $48$ is forbidden for they are all antisymmetrical in the $x$ direction.
Thus, there are only three intraband peaks and two interband peaks in Fig. \ref{fig:Fig5}(g).
Moreover, in the evolution from Fig. \ref{fig:Fig5}(g) to Fig. \ref{fig:Fig5}(c), two intraband peaks disappear due to too little overlap between the eigenstates [Fig. \ref{fig:Fig3}(b)].

In the $y$-polarized fine FIR spectrum [Fig. \ref{fig:Fig5}(h)], the intraband transitions are forbidden.
Moreover, Fig. \ref{fig:Fig5}(h) shows that two interband peaks in Fig. \ref{fig:Fig5}(g) are also forbidden.
In other words, all intraband transitions and some interband transitions are forbidden when the absorbed light is polarized along the direction of the electric field, which is similar to the finding in the previous section.
In the evolution from Fig. \ref{fig:Fig5}(h) to Fig. \ref{fig:Fig5}(d), interband absorption disappears on $U>0$ due to too little overlap between the eigenstates.

Fig. \ref{fig:Fig4}(d) can be regarded as the combination of Fig. \ref{fig:Fig5}(g) and (h).
Fig. \ref{fig:Fig4}(b) can be regarded as the combination of Fig. \ref{fig:Fig5}(c) and (d).
Thus, with $EF2$, all intraband transitions are $x$ polarized and all interband transitions are either $x$ polarized or $y$ polarized in the unpolarized FIR [Fig. \ref{fig:Fig4}(b) and (d)].
These polarized transitions are related to the parity selection rule as well as to the electric-field-direction-polarization selection rule.

\section{Summary}

In this paper, we have investigated the effects of $x$- and $y$-direction in-plane uniform electric fields on the FIR spectra of a single-electron triangular zigzag graphene quantum dot.
We have presented the energy spectra and the distribution of wave functions to analyse the FIR spectra.
The orthogonal zero-energy eigenstates are arranged along to the direction of the external field.
The remarkable result is that all intraband transitions and some interband transitions are forbidden when the absorbed light is polarized along the direction of the electric field.
With $x$-direction electric field, all intraband absorption is $y$ polarized due to the electric-field-direction-polarization selection rule.
Moreover, with $y$-direction electric field, all absorption is either $x$ or $y$ polarized due to the parity selection rule as well as to the electric-field-direction-polarization selection rule.
Our calculation shows that the formation of the FIR spectra is co-decided by the polarization selection rule and the overlap between the eigenstates.
These findings suggest that special attention should be paid to the property of polarization when designing TGQD optoelectronic devices.
Our findings may be useful for the application of GQDs to electronic and optoelectronic devices.
\section*{Acknowledgements}
This work is supported by the National Natural Science Foundation of China (Grant Nos. 11604183 and 11674197), and a Project of Shandong Province Higher Educational Science and Technology Program (Grant No. J16LJ09).

\end{document}